\documentclass[
 aps, pra,
 amsmath,amssymb,
 11pt,
 final,
tightenlines,
 twoside,
 twocolumn,
 floatfix,
nofootinbib,
 superscriptaddress,
showkeys,
 ]
{revtex4}
\usepackage{booktabs}
\usepackage{upgreek}
\usepackage{url}
\usepackage{natbib}
\usepackage{appendix}
\usepackage[T1]{fontenc}
\usepackage[utf8x]{inputenc}
\usepackage[english]{babel}
\usepackage{graphicx}
\usepackage{dcolumn}
\usepackage{bm}
\usepackage{longtable}
\input{maik.rty}

\setcitestyle{authoryear,round}
\def\saoname{Special Astrophysical Observatory,  Russian Academy of Sciences,
              Nizhnii Arkhyz, 369167 Russia}

\def\squareforqed{\hbox{\rlap{$\sqcap$}$\sqcup$}}

\def\sq{\ifmmode\squareforqed\else{\unskip\nobreak\hfil
\penalty50\hskip1em\null\nobreak\hfil\squareforqed
\parfillskip=0pt\finalhyphendemerits=0\endgraf}\fi}

\def\arcmin{\hbox{$^\prime$}}

\def\arcsec{\hbox{$^{\prime\prime}$}}

\def\utw{\smash{\rlap{\lower5pt\hbox{$\sim$}}}}

\def\udtw{\smash{\rlap{\lower6pt\hbox{$\approx$}}}}

\def\diameter{{\ifmmode\mathchoice
{\ooalign{\hfil\hbox{$\displaystyle/$}\hfil\crcr
{\hbox{$\displaystyle\mathchar"20D$}}}}
{\ooalign{\hfil\hbox{$\textstyle/$}\hfil\crcr
{\hbox{$\textstyle\mathchar"20D$}}}}
{\ooalign{\hfil\hbox{$\scriptstyle/$}\hfil\crcr
{\hbox{$\scriptstyle\mathchar"20D$}}}}
{\ooalign{\hfil\hbox{$\scriptscriptstyle/$}\hfil\crcr
{\hbox{$\scriptscriptstyle\mathchar"20D$}}}}
\else{\ooalign{\hfil/\hfil\crcr\mathhexbox20D}}%
\fi}}

 
\begin{document}
 
\selectlanguage{english}
 

 
 \title{A search for new dwarf galaxies outside the nearby groups}
 \author{\firstname{I.D.}~\surname{Karachentsev}}
  \email{ikar@sao.ru}
  \affiliation\saoname
  
  \author{\firstname{V.E.}~\surname{Karachentseva}}
    \affiliation{Main Astronomical Observatory, National Academy of Sciences of Ukraine, Kiev 03143, Ukraine}
  
  \author{\firstname{S.S.}~\surname{Kaisin}}
    \affiliation\saoname
  
  \author{\firstname{E.I.}~\surname{Kaisina}}
  \affiliation\saoname

\begin{abstract}  
We undertook a search for new nearby dwarf galaxies outside the known groups in the Local Volume using the data on DESI Legacy Imaging Surveys. In a wide sky area of $\sim$5000 square degrees directed toward the Local Void, we found only 12 candidates to nearby low mass galaxies. Almost all of them are classified as irregular or transition type dwarfs. Additionally, we examined areas of the sky exposed with the Hyper Suprime Camera of the Subaru telescope ($\sim$700 square degrees) and found nine more candidates to nearby dwarfs. Finally, nine candidates to the Local Volume were selected by us from the Zaritsky's SMUDG catalog that contains 7070 ultra-diffuse objects automatically detected in the whole area of the DESI surveys. We estimated a fraction of quiescent dSph galaxies in the general cosmic field to be less than 10 percent.
\end{abstract}

\maketitle

\section{Introduction} 
According to modern ideas, the spatial distribution of galaxies is similar to a cosmic ''web'', where particles (galaxies) are concentrated in filaments and walls that encompass large cosmic voids.  Groups and clusters form at the intersection of filaments and walls [1, 2]. Catalogs of groups and clusters in the nearby universe show that about half of all galaxies are part of virialized systems of varying populations, and about 15\% more galaxies are located in larger collapsing regions around groups, not participating in the global cosmological expansion [3--5]. Thus, about one-third of galaxies are located in the general metagalactic field and do not noticeably interact with each other.

General field galaxies are of particular interest because the star formation history in them follows a different scenario than the evolution of members of virialized systems. The most suitable object to study these differences is the Local Volume (LV) with a radius of $\sim$10~Mpc around the Milky Way, where observational data on galaxies are the most complete. By now, the number of known galaxies in this volume has approached a thousand [6] and continues to increase due to new surveys of the sky in optical and radio bands.

Most searches for new close dwarf galaxies have been undertaken in the region of known close groups [7--10].  This approach led to increase the statistics of the ''test particles'' used to determine the virial mass of the groups, thus making a more complete picture of the dark matter distribution in the LV.

Searches for dwarf galaxies not bound to nearby groups have been repeatedly undertaken [11--18].  A summary of observational data on the LV galaxies is presented in the Updated Nearby Galaxy Catalog (=UNGC, [6]) and the Local Volume Galaxy Database [19], the latest updated version of which is available at http://www.sao.ru/lv/lvgdb.

An obvious advantage of searching for dwarf galaxies in groups and clusters is the ability to assign to new objects a likely distance estimate that coincides with the group mean distance. In the general field dwarfs, their distance and total luminosity remain uncertain until the radial velocity is measured or structural features (e.g. globular clusters) suitable as distance indicators are found. In the LV, the best method is to measure the distance by the luminosity of the tip of red giant branch (TRGB).  This method, used in the Hubble Space Telescope (HST) observations, can measure the distance for any type of galaxy in the LV with an accuracy of $\sim$5\% or $\sim$(0.3--0.5) Mpc, allowing us to confidently separate field galaxies against group members.

\section{Searches for new dwarf galaxies in the Local Volume}
\subsection{The Local Void region}

\renewcommand{\baselinestretch}{0.5}
\begin{table*}
\caption{LV candidates outside the neighboring groups} 
\label{table1}
\begin{tabular}{l|c|c|c|l|l|c|c|c|c|c} \hline
Name    &	RA (2000.0) DEC  &     	$a'$ & $b/a$ &Type&$m_{\rm FUV}$& 	$m_{\rm NUV}$&  $B$ &	$g$&	$r$&	Notes\\ \hline
        &    h m s \,\,  deg\, $\arcmin \,\arcsec$ &	arcmin	&		& & mag&	mag&	mag&	mag&	mag& \\	   \hline
    (1)	    &       (2)  &(3)     	&(4)	  &   (5)	&  (6)	  &    (7)	&     (8)	&      (9)	&    (10)&	    (11)  \\  \hline
SMDG0956+82	&09 56 13.0 +82 53 24	&0.70	&0.60	&Irr&	19.64	&19.33&	18.08	&17.75	&17.42	&(1)  \\
Dw1155+78  	&11 55 54.2 +78 04 44	&0.58	&0.73	&Tr	&  $>$23.5	 &  ---  &   19.61	&19.18	&18.54	&---\\
Dw1234+76	&12 34 23.3 +76 43 34	&0.96	&0.58	&Irr&	21.76	&21.55&	17.9	&17.58	&-	   & (1) \\
Dw1339+39   &13 39 45.1 +39 08 09	&0.60	&0.66	&Irr&	18.65	&18.38&	17.68	&17.45	&17.51	&(2)\\
SMDG1349+04	&13 49 16.8 +04 49 05	&0.72	&0.73	&Tr	&  $>$23.5	2&1.88&	18.88	&18.53	&18.16	&---\\
Dw1459+44	&14 59 38.4 +44 40 23	&1.43	&0.95	&BCD&	17.02	&16.91&	17.07	&16.8	&16.78	&(3)\\
UGC9893sat	&15 32 51.4 +46 25 55	&0.71	&0.79	&Irr&	$>$23.5	&22.05&	20.63	&20.22	&19.62	&(4)\\
Dw1533+67	&15 33 28.1 +67 45 29	&0.83	&0.46	&Irr&	19.36	&19.06&	17.89	&17.59	&17.35	&--- \\
Dw1559+46	&15 59 02.6 +46 23 40	&0.60	&0.88	&Irr&  ---       & ---      &   17.19	&16.90	&16.70	&(1)\\
Dw1645+46	&16 45 48.5 +46 47 24	&0.74	&0.67	&Irr&	19.71	&19.48&	17.93	&17.63	&17.40	&(1) \\
Dw1709+74	&17 09 45.6 +74 10 44	&1.39	&0.81	&Tr	&   21.35	&20.79&	17.46	&17.04	&16.43	&(5)\\
Dw1735+57	&17 35 34.6 +57 48 47	&0.69	&0.66	&Irr&	18.35	&18.15&	17.07	&16.74	&16.40	&--- \\ \hline

\multicolumn{11}{l}{Notes: (1)~---  granulated; (2)~--- near a bright star; the object is located at 16$\arcmin$ to West from UGC 9660 } \\
\multicolumn{11}{l}{that has $V_h = 608$ km/s and $D$ = 10.76 Mpc via TRGB; (3)~--- a compact dwarf with an extended halo,} \\
\multicolumn{11}{l}{$V_h = 683$ km/s; (4)~--- the object has a very low surface brightness; it is only 2$\arcmin$ SW from UGC9893;}\\
 \multicolumn{11}{l}{(5)~--- probably a distant dwarf.} 
\end{tabular}
\end{table*}
\renewcommand{\baselinestretch}{1.0}

\begin{figure*}[hbt]
 \includegraphics[scale=0.5]{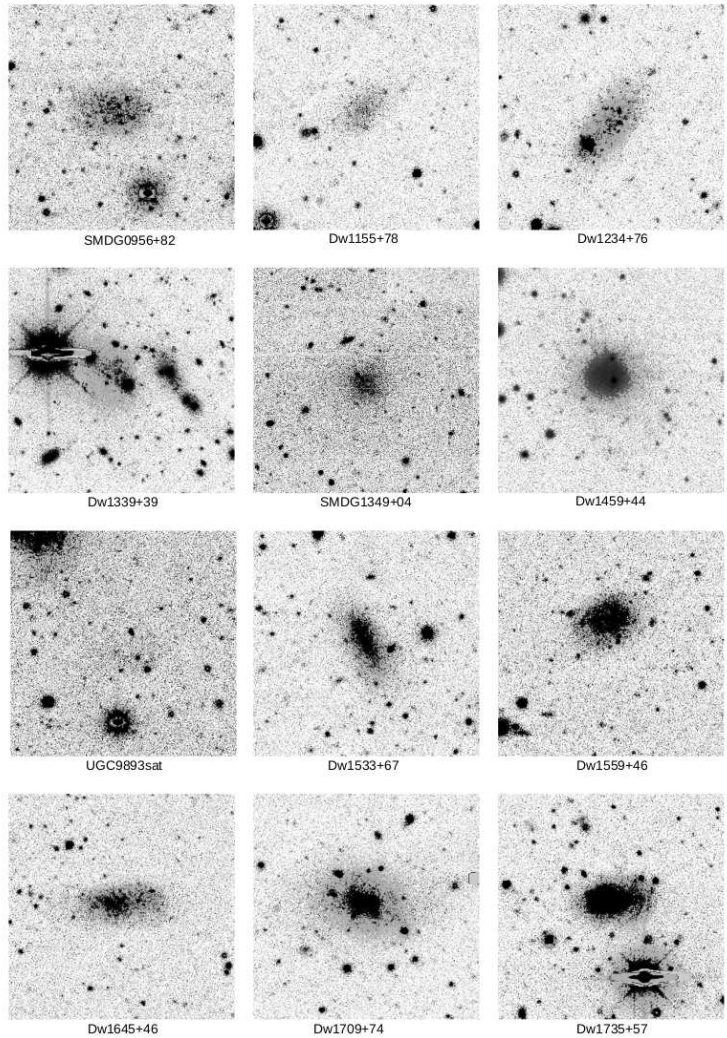}
 \caption{Images of 12 new dwarf galaxies from DESI Legacy Imaging Surveys found in the Local Void region. Each image size is $2\arcmin\times 2\arcmin$. North is to the top and East is to the left.}
 \end{figure*}

The publication of DESI Legacy Imaging Surveys [20] has provided an opportunity to discover new LV dwarf candidates over an area of $\sim$14000 square degrees, covering about one third of the entire sky. We have already used this survey to search for dwarf satellites around 46 relatively massive LV galaxies [10]. This resulted in the discovery of 67 satellite candidates for these galaxies, half of which turned out to be low surface brightness spheroidal dwarfs (dSph) with no evidence of star formation. To search for new dwarf galaxies in the local total field, we chose a wide region in the direction of the Local Void, whose center according to [21] has approximate equatorial coordinates RA $\simeq18^h$, Dec. $\simeq 0^{\circ}$. Our survey area covered the sky region RA $>13.5^h$, Dec.~$>-5^{\circ}$, the eastern side of which was bounded by the Milky Way band. Additionally, we inspected the polar cap with Dec.~$>+60^{\circ}$, avoiding the virial area of the known nearby group around M81. In the inspected direction, there is another void Draco-Cepheus [22] with its center at a distance of 14~Mpc. A panorama of the large-scale matter distribution in this direction is shown by [2]. In total, the area we inspected outside the nearby groups amounts to about 5000 square degrees or 1/3 of the DESI Legacy Imaging Surveys.  In this huge area, we detected only 12 candidates for new LV dwarf galaxies. Their images are shown as a mosaic in Fig. 1, where the size of each image corresponds to 2$\arcmin$.

The list of detected objects is given in Table 1, whose columns contain: (1)~--- galaxy name; (2)~--- equatorial coordinates in h, m, s  and deg,~$\arcmin,~\arcsec$; (3)~--- a maximum apparent galaxy  diameter in arc minutes; (4)~---  an apparent axial ratio; (5)~---  morphological type: irregular (Irr), spheroidal  (dSph), or intermediate (transition) (Tr); (6,7)~---  apparent galaxy magnitudes in the $FUV$ and $NUV$ bands of the  GALEX Ultraviolet Sky Survey [23]; (8)~---  apparent magnitude of the galaxy in the  $B$-band; (9,10)~--- apparent magnitudes in the $g$- and $r$-bands; (11)~---  notes on the presence of the  galaxy's structure features and close surroundings.

The total apparent $g$- and $r$-magnitudes of galaxies were taken from the SMUDG Ultra-Diffuse Galaxy Catalog [17] or from the Legacy Survey DR9 photometric data [20].  For some low surface brightness galaxies with patchy structure, we have made $g$- and $r$-magnitude measurements for galaxy images from the DESI Legacy Imaging Surveys using the standard methods of processing in the MIDAS software package from the European Southern Observatory. A typical measurement error was about 0.10 mag. The total $B$-magnitude of galaxies was determined using the ratio $B = g +0.313(g-r) +0.227$ recommended by Lupton\footnote{https://www.sdss3.org/dr10/algorithms /sdssUBVRITransform.php\#Lupton2005}. The median apparent magnitude of the detected dwarf galaxies is $B\simeq18^m$, which at the far edge of the LV ($D$~=~10~Mpc) corresponds to an absolute magnitude of $M_B \simeq -12^m$.
All of these dwarf galaxies appear to be rather isolated objects located far away from other known members of the Local Volume. The only exception is a very low surface brightness dwarf, UGC9893sat, which is almost in contact with another peculiar dIrr galaxy UGC9893 = IZw115 = VV720 = KIG 686, which has a distance estimate of 10.2 Mpc based on the Tully-Fisher relation between a galaxy luminosity and its HI-line width. It is likely that these dwarf galaxies form a physical pair at a stage close to their merger.

\subsection{HSC survey areas} 
Among the DESI Legacy Imaging Surveys (http://www.legacysurvey.org/), there is a series of sky areas exposed on the 8.5-meter Subaru telescope Hyper Suprime  Camera (HSC) with a deeper limit and higher angular resolution in comparison with the Legacy  surveys. The combined area of them is about 700 square degrees. We inspected these regions of the sky and found 9 other LV dwarf galaxy candidates. Their images  are shown in Fig. 2, where each element of the mosaic is $2\arcmin$ in size. The data on these objects are compiled in Table 2, in which the column designations are the same as in Table 1. Some individual features of the galaxies are presented in notes to the table. Among these objects there are only two dwarf galaxies of spheroidal type. Of them, Dw0852-0210 may be a companion of the Sc galaxy UGC4640 having a heliocentric velocity $V_h =3308$ km/s, and Dw1229+0144 may be a companion of the Sbc galaxy NGC4536 with $V_h =1807$ km/s or be belong to peripheral members of the Virgo cluster.

 \renewcommand{\baselinestretch}{0.5}
 \begin{table*}
\caption{LV candidates from survey of HSC-fields in DESI Legacy Imaging Surveys} \label{table2}
\begin{tabular}{l|c|c|c|l|l|c|c|c|c|c}  \hline
Name    &	RA (2000.0) DEC  &     	$a'$ & $b/a$ &Type&$m_{\rm FUV}$& 	$m_{\rm NUV}$&  $B$ &	$g$&	$r$&	Notes\\ \hline
        &    h m s \,\,  deg\, $\arcmin\, \arcsec$ &	arcmin	&		& & mag&	mag&	mag&	mag&	mag& \\	   \hline
    (1)	    &       (2)  &(3)     	&(4)	  &   (5)	&  (6)	  &    (7)	&     (8)	&      (9)	&    (10)&	    (11)  \\  \hline

Dw0010-0112&	00 10 26.4 $-$01 12 18&	0.58&	0.61&	Tr	&$>$23.5&	---	  &  20.58&	20.2  &	19.7	&(1) \\
SMDG0220-00&	02 20 48.2 $-$00 27 54&	0.97&	0.83&	Tr	&22.79&	20.9	&18.04&	17.65	&17.13	&(2) \\
SMDG0223-02&	02 23 18.7 $-$02 03 25&	0.66&	0.67&	BCD	&23.5:&	---	  &  18.68&	18.32	&17.93	&(3) \\
Dw0852-0210&	08 52 35.5 $-$02 10 41&	0.73&	0.97&	Sph	&$>$23.5&	 ---   	&20.01&	19.55	&18.8	&(4) \\
Dw1229+0144&	12 29 22.1 $+$01 44 17&	1.11&	0.68&	Sph	&23.49&	21.53	&18.66&	18.22	&17.53	&(5)\\
Dw1235-0215&	12 35 53.0 $-$02 15 54&	0.64&	0.82&	Irr	&20.13&	20.04	&19.02&	18.7	1&8.39	&(6)\\
SMDG1245+01&	12 45 52.3 $+$01 54 22&	0.85&	0.56&	Sph	&$>$23.5&	 ---&  18.01&	17.61	&17.05	&(7)\\
SMDG1256-00&	12 56 35.8 $-$00 31 44&	0.74&	0.67&	Irr &20.28&	19.65	&18.14&	17.8	1&7.44	&(8)\\
Dw1416-0212&	14 16 38.9 $-$02 12 04&	0.81&	0.89&	Im	&20.47&	19.62	&18.17&	17.83	&17.46	&(9)\\ \hline
  
  \multicolumn{11}{l}{Notes: (1)~--- there are some cirrus in its vicinity; (2)~--- $V_h = 4400\pm6000$ km/s from [24];}\\
  \multicolumn{11}{l}{(3)~--- it is located at 20$\arcmin$ to NW from UGC 1862 that has $V_h = 1383$ km/s; (4)~--- very low }\\
  \multicolumn{11}{l}{surface brightness; above the bright star; at 12$\arcmin$ to W there is a galaxy UGC 4640 with}\\
  \multicolumn{11}{l}{ $V_h = 3309$ km/s; (5)~--- in contact with a bright star, may be a member of Virgo cluster;}\\
  \multicolumn{11}{l}{(6)~--- granulated, a probable satellite of the Sombrero galaxy; (7)~--- granulated; at 40$\arcmin$ to W}\\
  \multicolumn{11}{l}{there is a bright peculiar SB galaxy NGC 4643 with 3 tight dSphs and $V_h = 1333$ km/s}\\
  \multicolumn{11}{l}{(the Virgo cluster members?); (8)~--- probably a peripheric member of the Virgo cluster;}\\
  \multicolumn{11}{l}{(9)~--- with a faint asymmetric envelope; not detected in HIPASS.}\end{tabular}
  \end{table*}

  \renewcommand{\baselinestretch}{1.0}

\subsection{Candidate dwarf galaxies of the LV from the SMUDG catalog} 
The regions of the sky, where our search for new Local Volume candidates was conducted, cover a significant part, but not the entire DESI Legacy Imaging Surveys area.  Therefore, we used data from the SMUDG catalog (Systematically Measuring Ultra-Diffuse Galaxies, [24]) to continue our search. The latest full version of this catalog [17] contains 7070 low surface brightness objects automatically extracted from DR9 Legacy Survey images, with the results of surface photometry of these objects presented in the $g$- and $r$-bands. As shown by our analysis, along with dwarf galaxies, the catalog includes a number of fragments of interstellar cirrus. Visual inspection of the SMUDG objects allowed us to identify 9 dwarf galaxies that can be located at a distance $D\leq12$ Mpc. The list of them is summarized in Table 3. The content of its columns is analogous to the corresponding data in Tables 1 and 2. The images of these galaxies are reproduced in Fig. 3 in the same manner as in the previous figures.

Among the 9 galaxies, one, KK220=AGC232003 has a radial velocity measurement,
$V_{\rm hel} = 776\pm6$ km/s, which allows us to estimate the galaxy distance $D$=11.1~Mpc based on the Numerical Action Method model (NAM, [25]).  According to the table annotations, two dSph galaxies are well-isolated objects, while the other two dSphs may be members of more distant systems outside the LV.
   
   \begin{figure*}[hbt]
 \includegraphics[scale=0.5]{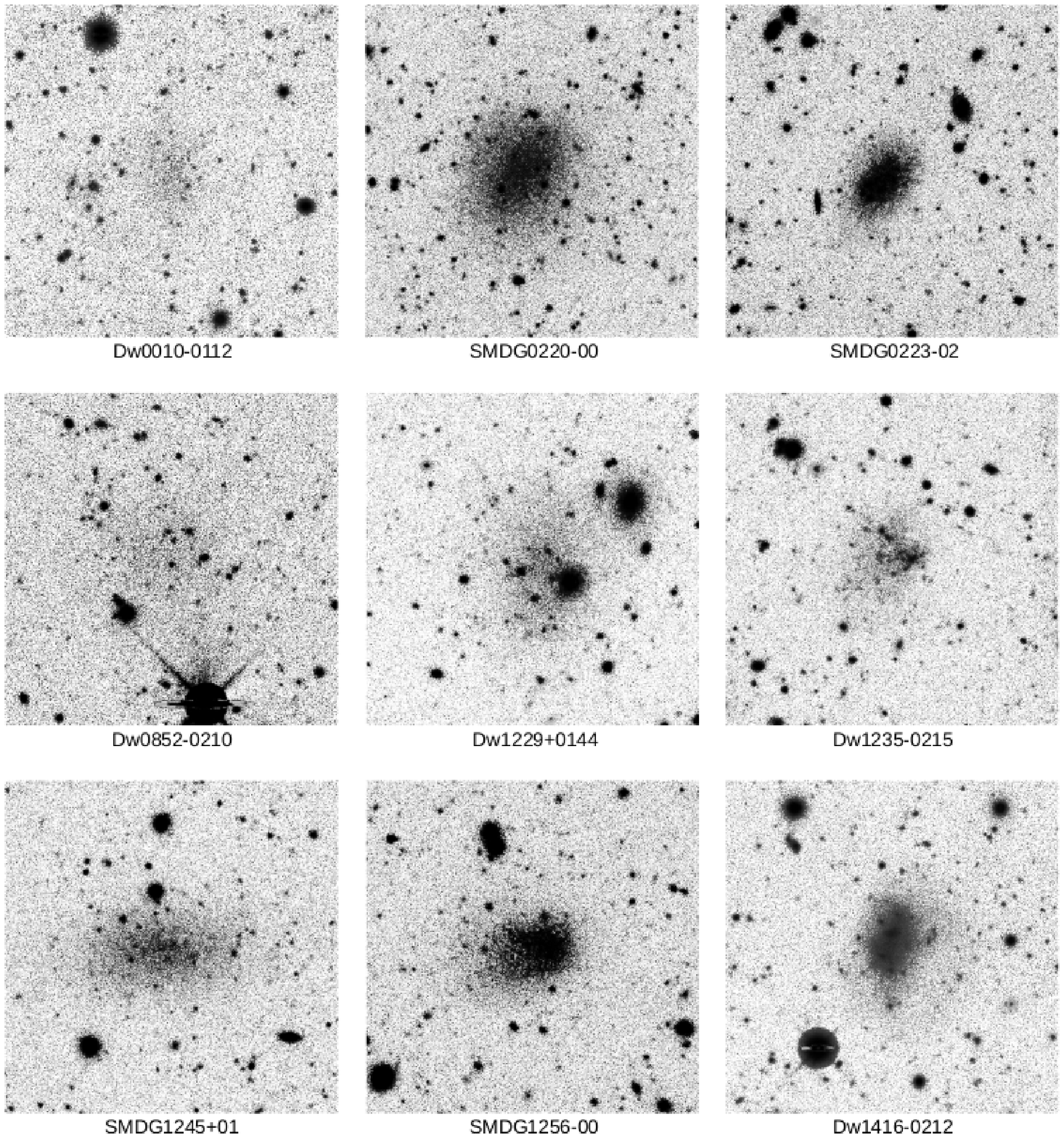}
 \caption{Images of 9 new dwarf galaxies found in areas of the Subaru telescope Hyper SuprimeCamera survey. Each image size is $2\arcmin\times 2\arcmin$. North is to the top and East is to the left.}
 \end{figure*}

   \renewcommand{\baselinestretch}{0.5}   
      \begin{table*} \label{table3}
\caption{LV candidates from Zaritsky+2023 catalog of ultra-diffuse galaxies.}
\begin{tabular}{l|c|c|c|l|l|c|c|c|c|c}  \hline
Name    &	RA (2000.0) DEC  &     	$a'$ & $b/a$ &Type&$m_{\rm FUV}$& 	$m_{\rm NUV}$&  $B$ &	$g$&	$r$&	Notes\\ \hline
        &    h m s \,\,  deg\, $\arcmin\, \arcsec$ &	arcmin	&		& & mag&	mag&	mag&	mag&	mag& \\	   \hline
    (1)	    &       (2)  &(3)     	&(4)	  &   (5)	&  (6)	  &    (7)	&     (8)	&      (9)	&    (10)&	    (11)  \\  \hline
SMDG0740+40	  &   07 40 23.0 $+$40 32 56&	0.48&	0.63&	Irr&	21.46&	---	&19.27	&18.87	&18.30	&(1) \\
SMDG1017-05   &   10 17 16.3 $-$05 29 53&	0.69&	0.72&	Sph&	$>$23.5&	---	&19.17	&18.75	&18.13	&(2)  \\
SMDG1103+60	  &   11 03 56.4 $+$60 29 53&	0.43&	0.64&	Irr&	  ---&---	    &19.48	&19.16	&18.88	&--- \\
SMDG1110+18   &   11 10 55.0 $+$18 58 52&	0.63&	0.83&	Sph&	$>$23.5&	 ---	&18.53	&18.14	&17.61	&(2) \\
SMDG1221+28   &   12 21 49.7 $+$28 31 12&	0.47&	0.74&	Sph&	$>$23.5&	---	&19.35	&18.95	&18.39	&(3) \\
SMDG1241+35	  &   12 41 11.0 $+$35 11 46&	0.73&	0.92&	Irr&	21.8 & 21.08&	18.90&	18.53&18.09	&--- \\
SMDG1250+07   &   12 50 26.9 $+$07 44 35&	1.33&	0.54&	Sph&	$>$23.5&	 ---	&17.99	&17.53	&16.78	&(4)  \\
SMDG1345+33	  &   13 45 11.0 $+$33 11 31&	0.43&	0.86&	Irr&    $>$23.5& 22.47&	19.10&	18.71&	18.2&	(5)\\
KK220=A232003 &	  13 47 36.5 $+$33 12 22&	1.11&	0.93&	Im	&   18.92& 18.07&	16.3&8	16.02&	15.59&	(6)\\ \hline

 \multicolumn{11}{l}{Notes: (1)~--- at 30$\arcmin$ to S there is DDO 46 with D = 10.38 Mpc via TRGB;(2)~--- an isolated dSph;}\\
\multicolumn{11}{l}{(3)~--- may be a satellite of NGC 4251 with $V_h$ = 1066 km/s or NGC 4278 with $V_h$ = 649 km/s and}\\
\multicolumn{11}{l}{$D$ = 16.07 Mpc  via sbf; (4)~--- granulated; in the Virgo cluster?; (5)~--- a probable pair with the KK 220;}\\
\multicolumn{11}{l}{(6)~--- It has $V_h$ = 776 km/s, $W_{50}$ = 19 km/s and $m_{21}$ = 17.1 mag, $D$ = 11.1 Mpc via NAM method.} \end{tabular}

  \end{table*}
  \renewcommand{\baselinestretch}{1.0}

\section{The problem of distance estimation for dSph galaxies}  
Dwarf dIrr- and BCD-galaxies rich in gas and with active star formation are easy targets for measuring their radial velocity both in the optical and radio ranges. The situation with determination of the radial velocity in dwarfs of spheroidal and transitional (Tr) types is much more complicated. The absence of star formation sites as well as appreciable amount of neutral hydrogen makes it almost impossible to measure the velocity of a galaxy. In nearby dSph objects, it is sometimes possible to determine the optical velocity in the presence of a globular cluster or planetary nebula in the galaxy.
  $$SB = (1/3) \times M_B + const,$$
  
    \begin{figure*}
 \includegraphics[scale=0.5]{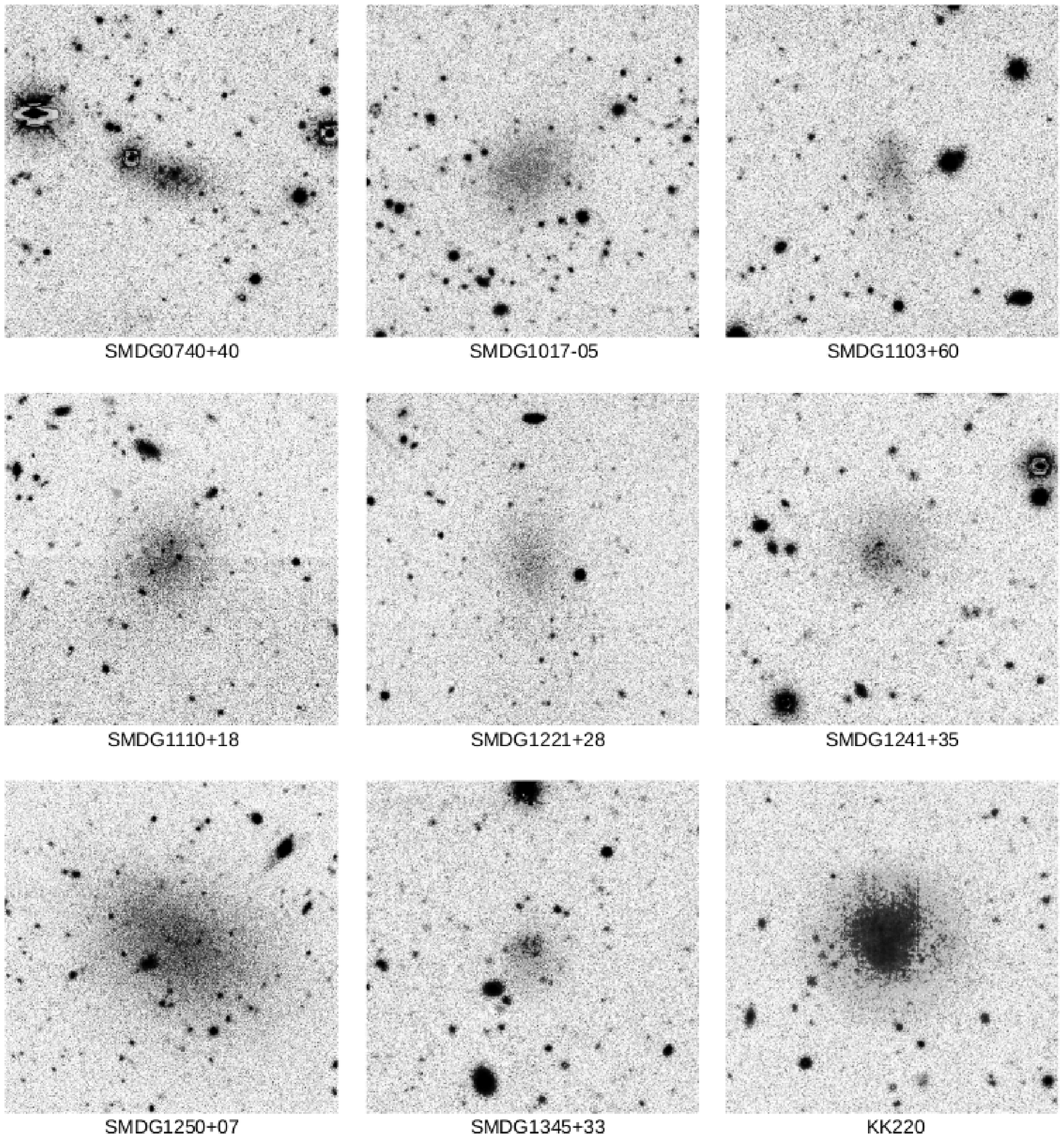}
 \caption{Images of 9 the Local Volume galaxy candidates from Zaritsky's SMUDG catalog.  Each image size is $2\arcmin\times 2\arcmin$. North is to the top and East is to the left.}
 \end{figure*}
  
There is a dependence between the absolute magnitude of dSph-galaxy and its average
surface brightness which reflects the approximate constancy of the mean spatial stellar density in galaxies [26]. This relation can be used to estimate roughly the distance to a Sph or Tr type galaxy. However, a scatter in the diagram {M vs. SB} is quite large, which produces significant errors in the distance estimates. For example, about a dozen dSph galaxies were found in the nearby ($D$=6.9 Mpc) group around the luminous Sc galaxy M101, which were thought to be members of this group [27]. However, the HST observations of these galaxies have disproved this naive assumption.  These low surface brightness galaxies turned out to be members of more distant groups at distances of $D\sim$~28 Mpc. A similar situation happened with dSph galaxies: KKR08, KDG218, KDG229, where the HST observations did not confirm their belonging to the LV.

Looking through large portions of the sky in Legacy Imaging Surveys, we have found many cases where very low surface brightness tidal structures are seen around a bright host galaxy.  They are usually similar in size and brightness contrast to isolated ultra-diffuse dwarf galaxies. The tidal nature of such semi-detached dwarfs is quite obvious. In Table 4, we present a few examples where it is difficult to make the distinction between clumps in tidal structures and detached dSph galaxies. The table contains the name of the bright galaxy, its equatorial coordinates in degrees, the galaxy redshift, and the characteristic angular size of the field in which neighboring dwarfs and tidal structures are visible.
    
Two fields around the galaxies NGC4643 and NGC5557 are presented in Fig. 4. The low- contrast structures on them are marked with dashed ellipses. In the case of NGC4643, the dSph galaxy SMUDG1245+01 shown in Table 2 is located $\sim40\arcmin$. East of NGC4643, i.e. outside the image frame. The three noted clumps (or detached spheroidal dwarfs?) bear a strong resemblance to SMUDG1245+01. The tidal structures around NGC5557 are barely distinguishable from usual ultra-diffuse dwarf galaxies. These faint objects near NGC5557 are also well seen in the deep image obtained by [28]. Other examples of similar structures found in Legacy Imaging Surveys have been demonstrated recently by [29].
  
  \begin{figure*}
 \includegraphics[scale=0.5]{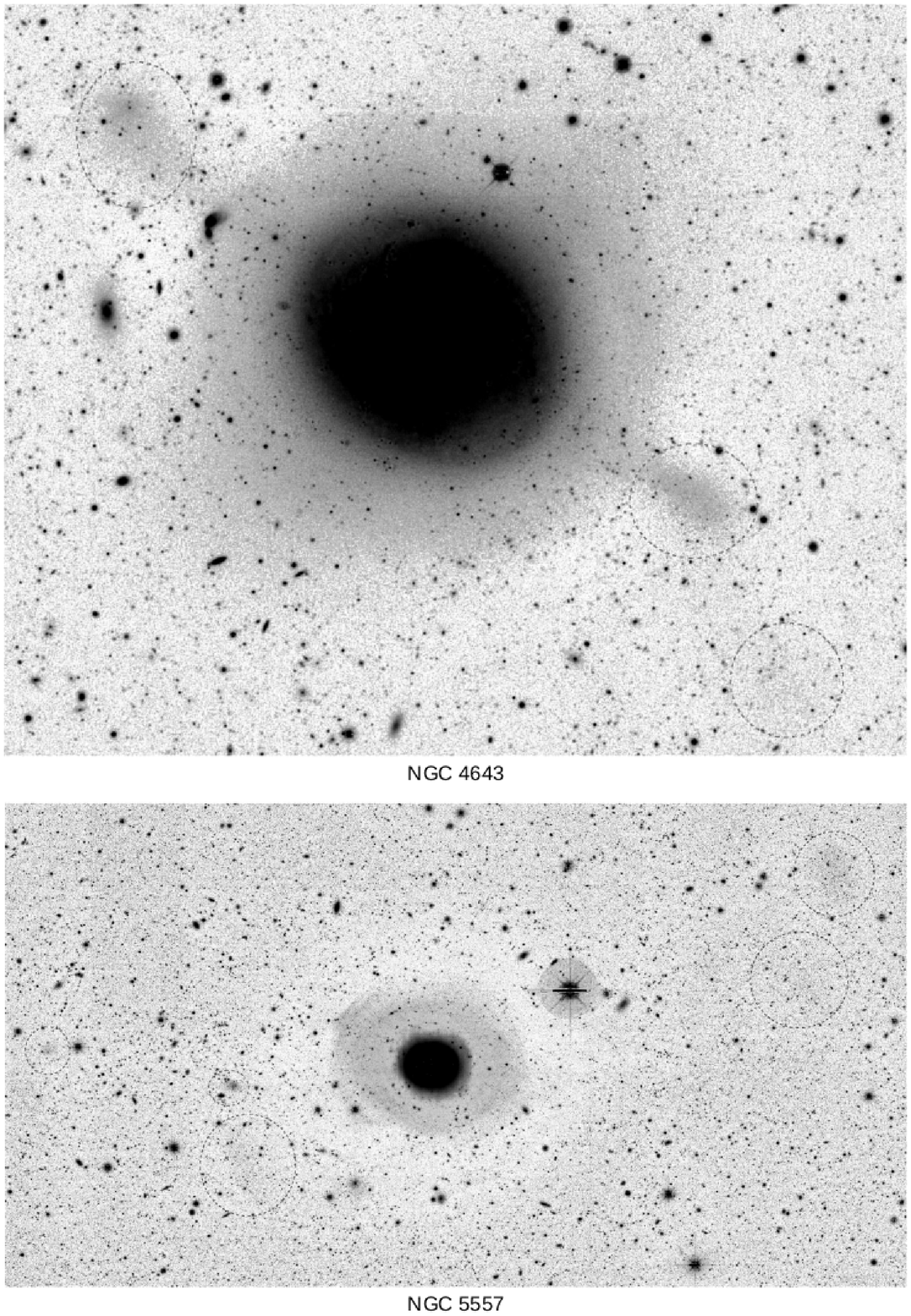}
 \caption{Images of the nearby bright galaxies: NGC 4643 (upper panel) and NGC 5557 (bottom panel) with low surface brightness objects around them indicated by dashed ellipses. The size of the images are $12\arcmin\times 10\arcmin$ for NGC 4643 and $28\arcmin\times 15\arcmin$ for NGC 5557. }
  \end{figure*}
 
\renewcommand{\baselinestretch}{0.5}
  
  \begin{table}
\caption{Borderline cases of semi-separated tidal dwarf galaxies.}
\begin{tabular}{l|c|c|c}  \hline
Name	    & RA (2000.0) DEC& 	 Z	   & Field size\\  \hline
	        & degr      degr	&	   & arcmin  \\ \hline
NGC 4643	&190.834 +01.978	&0.0044&	12 \\
PGC 049168	&207.751 +36.954	&0.0174&	6  \\
NGC 5363	&209.030 +05.255	&0.0038&	20 \\
NGC 5557  	&214.607 +36.494	&0.0107&	28 \\
UGC 9386	&218.720 +40.748	&0.0190	&   6   \\
NGC 5754	&221.332 +38.731	&0.0147&	8  \\
IC 1069	    &222.694 +54.411	&0.0376&	6  \\
PGC 055442  &233.454 +21.136	&0.0234&	6 \\
NGC 5971  	&233.904 +56.462	&0.0112&	6 \\
NGC 6020	&239.284 +22.405	&0.0144&	8 \\
PGC 2308494	&248.470 +47.998	&0.0354&	6 \\ \hline
\end{tabular}
\end{table}
\renewcommand{\baselinestretch}{1.0}

\section{Discussion and conclusion} 
Our experience with the search for new nearby dwarf galaxies in regions of the sky outside of close groups has led us to conclude that the efficiency of this hunt is low. For instance, our search for new members in 46 LV groups undertaken with DESI Legacy Imaging Surveys found 67 probable members of these groups over a total area of 280 square degrees [10]. Performing a survey of the broad vicinity of the Local Void on the same material and in the same manner, we found only 12 Local Volume candidates over an area of about 5000 square degrees. The effectiveness of searches in the general field, outside the groups, turns out to be two orders of magnitude lower than that in the groups themselves. This is consistent with the well-known fact that the slope of the galaxy luminosity function at its faint end is much smaller for field galaxies than for members of groups and clusters [30].

The population of the general field and cosmic voids is dominated by dwarf galaxies richer in gas and with a more active star formation rate [31]. According to our data presented in Tables 1--3, quiescent dSph galaxies account for a quarter of the total.  Some of them may turn out to be members of the nearby Virgo cluster.  However, the examples of dSph galaxies SMUDG1017-05 and SMUDG1110+18 suggest that a small percentage of dwarf spheroidal galaxies may be well-isolated objects. In the vicinity of the Local Group, the dSph galaxies KKR25 and KKs3 locate at a distance of $\sim$2 Mpc from the Milky Way, i.e. well beyond the virial radius of the Local Group [32, 33]. Investigating the distribution of dwarf satellites around massive Milky Way-type LV galaxies, [34] found that the fraction of dSph dwarfs among satellites with $K$-band luminosity $L/L_{\odot} >10^7$ decreases with the projected distance from the host galaxy Rp according to the exponential law

 $q(R_p) = 0.75 \times exp(-R_p /350\,\, {\rm kpc}).$
 
Then outside the radius of the zero-velocity sphere of a typical nearby group ($R_0\simeq1.0$ Mpc), i.e., in the general expanding field, the relative number of quiescent dSph dwarfs is about 5\%.  The presence of two definitely isolated dSph galaxies among the 28 ones found in our survey agrees well with the above estimate.

It is interesting to note, that Park et al, [35] performed measurements of satellite radial velocities around isolated massive early-type galaxies and found a lack of faint satellites with absolute magnitudes $M_r$ fainter than --14.0 mag. This effect, if confirmed on large statistics, can be caused by a peculiarity of the dynamical evolution of isolated elliptical galaxies.

Finally, we note a planned photometric survey of the sky with small-aperture telescopes of The Dragonfly Telephoto Array [36]. The authors expect to find a lot of new dwarf galaxies in the general field. They estimate that the number of new dwarfs in the LV up to the surface brightness limit of $\sim$30 mag $\times$ arcsec$^2$ could reach $\sim$0.1 per square degree or N $\sim10^3$ across the entire sky.

Acknowledgements. In the work we use the DESI Legacy Imaging Surveys and the Galaxy Evolution Explorer Survey. This work is supported by the grant 075-15-2022-262 (13.MNPMU.21.0003) of the Ministry of Science and Higher Education of the Russian Federation.
\clearpage

REFERENCES

1.~Wilding G., Nevenzeel K., van de Weygaert R., et al., MNRAS, 507, 2968,  2021.

2.~Courtois H.M., van de Weygaert R., Aubert M., et al., A \& A, 673A, 38, 2023.

3.~Vennik J.,  Tartu Astrofuusika Observatorium Treated, 73, 1, 1984.

4.~Makarov D.I., Karachentsev I.D., MNRAS, 412, 2498 , 2011.

5.~Kourkchi E., Tully R.B., ApJ, 843, 16,  2017.

6.~Karachentsev I.D., Makarov D.I.,  Kaisina E.I., AJ, 145, 101, 2013.

7.~Chiboucas K., Karachentsev I.D., Tully R.B., AJ, 137, 3009,  2009.

8.~Muller O., Rejkuba M., Pawlowski M.S., et al., A \& A, 629, A18,  2019.

9.~Carlsten, S.G., Greene, J.E., Beaton, R.L., ApJ, 933, 47,  2022.

10.~Karachentsev I.D., Kaisina E.I., Astr.Bull., 77, 411, 2022.\\
11.~van den Bergh S., A Catalog of Dwarf Galaxies, Publ. David Dunlap obs., 2, 147, 1959.

12.~Karachentseva V.E.,  Soobsch. Byurakan astrophys. Obs., 39, 61, 1968.

13.~Karachentseva V.E., Astron. Tsirk., 723, 1, 1972.

14.~Karachentseva V.E., Izvestia SAO, 5, 10, 1973.

15.~Karachentsev I.D., Karachentseva V.E., Suchkov A.A., Grebel E.K.,  A\&AS, 145, 415, 2000.

16.~Karachentseva V.E., Karachentsev I.D., A\&AS, 127, 409,  1998.

17.~Zaritsky D., Donnerstein R., Dey A., et al., ApJS, 267, 27, 2023.

18.~Paudel S., Yoon S.J., Yoo J., et al., ApJS, 265, 57, 2023.

19.~Kaisina E.I., Makarov D.I., Karachentsev I.D., et al., AstBul, 67, 115,  2012.

20.~Dey A., Schlegel D.J., Lang D., et al.,  AJ, 157, 168, 2019.

21.~Tully R.B., Nearby Galaxies Catalog, Cambridge University Press,  1988.

22.~Pustilnik S.A., Tepliakova A.L., Makarov D.I., MNRAS, 482, 4329,  2019.

23.~Martin D.C., Fanson J., Schiminovich D., et al., ApJ, 619, L1,  2005.

24.~Zaritsky D., Donnerstein R., Karunakaran A., et al., ApJS, 261, 11, 2022.

25.~Shaya E. L., Tully R. B., Hoffman Y.,  Pomarede D., ApJ, 850, 207,  2017.

26.~Karachentsev I.D., Karachentseva V.E., Huchtmeier W.K., Makarov D.I.,  AJ, 127, 2031, 2004.

27.~Merritt A., van Dokkum P., Abraham R.,  ApJ, 787, L37, 2014.

28.~Duc P.A., Paudel S., McDermid R.M., et al.,  MNRAS, 440, 1458, 2014.

29.~Martinez-Delgado D., Cooper A.P., Roman J., et al.,  A \& A, 671A, 141, 2023.

30.~Driver S. , De Propris R.,  ASpSci, 285, 175, 2003.

31.~Rodriguez-Medrano A.M., Paz D.J., Stasyszyn F.A., et al.,  MNRAS, 521, 916, 2023.

32.~Makarov D.I., Makarova L.N., Sharina M.E., et al.,  MNRAS, 425, 709, 2012.

33.~Karachentsev I.D., Kniazev A.Y., Sharina M.E.,  AN, 336, 707, 2015.

34.~Karachentsev I.,  Kashibadze O., Astr. Nachr., 342, 99, 2021.

35.~Park C., Hwang H.S., Park H., Lee J.C., NatAs, 2, 162,  2018.

36.~Danieli S., van Dokkum P., Conroy C.,  ApJ, 856, 69, 2018.
\end{document}